\documentclass[manuscript]{aastex}

\usepackage{graphicx}

\slugcomment{Accepted for Publication in Astronomical Journal}

\shorttitle{Dust in NGC 1512/1510}
\shortauthors{Liu et al.}

\begin{document}

%% LaTeX will automatically break titles if they run longer than
%% one line. However, you may use \\ to force a line break if
%% you desire.

\title{An Investigation of the Dust Content in the Galaxy pair \\
	NGC 1512/1510 from Near-Infrared to Millimeter Wavelengths}

%% Use \author, \affil, and the \and command to format
%% author and affiliation information.
%% Note that \email has replaced the old \authoremail command
%% from AASTeX v4.0. You can use \email to mark an email address
%% anywhere in the paper, not just in the front matter.
%% As in the title, use \\ to force line breaks.

\author{Guilin Liu\altaffilmark{1}, Daniela Calzetti\altaffilmark{1}, 
Min S. Yun\altaffilmark{1}, Grant W. Wilson\altaffilmark{1}}
\email{gliu@astro.umass.edu, calzetti@astro.umass.edu, 
myun@astro.umass.edu,\\ wilson@astro.umass.edu}

\author{Bruce T. Draine\altaffilmark{2}}
\email{draine@astro.princeton.edu}

\author{Kimberly Scott\altaffilmark{1}}
\email{kscott@astro.umass.edu}

\author{Jason Austermann\altaffilmark{1,3}}
\email{jaustermann@gmail.com}

\author{Thushara Perera\altaffilmark{1}}
\email{thusharaprr@gmail.com}

\author{David Hughes\altaffilmark{4}, Itziar Aretxaga\altaffilmark{4}}
\email{dhughes@inaoep.mx, itziar@inaoep.mx}

\author{Kotaro Kohno\altaffilmark{5,6}}
\email{kkohno@ioa.s.u-tokyo.ac.jp}

%\and
\author{Ryohei Kawabe\altaffilmark{7} and Hajime Ezawa\altaffilmark{7}}
\email{ryo.kawabe@nao.ac.jp, h.ezawa@nao.ac.jp}

\altaffiltext{1}{Astronomy Department, University of Massachusetts, Amherst, MA 01003, USA}
\altaffiltext{2}{Department of Astrophysical Sciences, Princeton, NJ 08544, USA}
\altaffiltext{3}{Center for Astrophysics and Space Astronomy, University of Colorado, Boulder, CO 80309, USA}
\altaffiltext{4}{Instituto Nacional de Astrof\'isica, \'Optica y Electr\'onica (INAOE), Aptdo. Postal 51 y 216, 72000 Puebla, Mexico}
\altaffiltext{5}{Institute of Astronomy, University of Tokyo, 2-21-1 Osawa, Mitaka, Tokyo 181-0015, Japan}
\altaffiltext{6}{Research Center for the Early Universe, University of Tokyo, 7-3-1 Hongo, Bunkyo, Tokyo 113-0033, Japan}
\altaffiltext{7}{Nobeyama Radio Observatory, National Astronomical Observatory of Japan, Minamimaki, Minamisaku, Nagano 384-1305, Japan}

\begin{abstract}
We combine new ASTE/AzTEC 1.1~mm maps of the galaxy pair NGC 1512/1510 with archival 
Spitzer IRAC and MIPS images covering the wavelength range 3.6--160~$\mu$m from the 
SINGS project. The availability of the 1.1~mm map enables us to measure the long--wavelength 
tail of the dust emission in each galaxy, and in sub--galactic regions in NGC 1512, and to 
derive accurate dust masses. The two galaxies form a pair consisting of a large, 
high--metallicity spiral (NGC 1512) and a low metallicity, blue compact dwarf (NGC 1510), 
which we use to compare similarities and contrast differences. Using the models of 
\citet{DraineLi07}, the derived total dust masses are (2.4$\pm$0.6)$\times$10$^7$~M$_{\odot}$ 
and (1.7$\pm$3.6)$\times$10$^5$~M$_{\odot}$ for NGC 1512 and NGC 1510, respectively. The 
derived ratio of dust mass to {\sc H i} gas mass for the galaxy pair, $\rm M_d/M_{HI}\sim$0.0034,
is much lower (by at least a factor of 3) than expected, as previously found by \citet{Draine07}. 
In contrast, regions within NGC 1512, specifically the central region and the arms, do not 
show such unusually low $\rm M_d/M_{HI}$ ratios; furthermore, the dust--to--gas ratio is within 
expectations for NGC 1510. These results suggest that a fraction of the {\sc H i} 
included in the determination of the $\rm M_d/M_{HI}$ ratio of the NGC 1512/NGC 1510 pair is 
not associated with the star forming disks/regions of either galaxy. Using the dust masses 
derived from the models of \citet{DraineLi07} as reference, we perform simple two--temperature 
modified--blackbody fits to the far--infrared/mm data of the two galaxies and the sub--regions of 
NGC 1512, in order to derive and compare the dust masses associated with warm and cool dust 
temperature components. As generally expected, the warm dust temperature of the low--metallicity, 
low--mass NGC 1510 ($\rm T_w$$\sim$36~K) is substantially higher than the corresponding warm 
temperature of the high--metallicity spiral NGC 1512 ($\rm T_w$$\sim$24~K). In both galaxies 
(albeit with a large uncertainty for NGC 1510), our fits indicate that a substantial fraction 
($>$93\%) of the total dust mass is in a cool dust component, with temperatures $\sim$14--16~K 
for NGC 1512 and $\sim$15--24~K for NGC 1510. This result is similar to what is determined for a 
few other nearby galaxies. In contrast, the warm dust component in the sub--galactic regions of  
NGC 1512 represents a much larger fraction of the total dust content, in agreement with the fact 
that all three regions have higher specific star formation rates than the average in the galaxy;
in the center, the warm dust represents about 40\% of the total, while in the arms the fractions 
are close to $\sim$20\%. 

\end{abstract}

%% Keywords should appear after the \end{abstract} command. The uncommented
%% example has been keyed in ApJ style. See the instructions to authors
%% for the journal to which you are submitting your paper to determine
%% what keyword punctuation is appropriate.

\keywords{galaxies: individual (NGC 1510, NGC 1512); galaxies: ISM; 
galaxies: starburst; (ISM:) dust, extinction}
%\keywords{globular clusters: general --- globular clusters: individual(NGC 6397,
%NGC 6624, NGC 7078, Terzan 8}

%% From the front matter, we move on to the body of the paper.
%% In the first two sections, notice the use of the natbib \citep
%% and \citet commands to identify citations.  The citations are
%% tied to the reference list via symbolic KEYs. The KEY corresponds
%% to the KEY in the \bibitem in the reference list below. We have
%% chosen the first three characters of the first author's name plus
%% the last two numeral of the year of publication as our KEY for
%% each reference.

%% Authors who wish to have the most important objects in their paper
%% linked in the electronic edition to a data center may do so by tagging
%% their objects with \objectname{} or \object{}.  Each macro takes the
%% object name as its required argument. The optional, square-bracket 
%% argument should be used in cases where the data center identification
%% differs from what is to be printed in the paper.  The text appearing 
%% in curly braces is what will appear in print in the published paper. 
%% If the object name is recognized by the data centers, it will be linked
%% in the electronic edition to the object data available at the data centers  

\section{INTRODUCTION}

Interstellar dust provides a source of attenuation of the light from galaxies,
but also a source of information on their metal content, and, possibly, a
tracer for the molecular gas content \citep{Draine07}. On average, about
50\% of the UV--optical light from galaxies is re-processed by their own dust
into the mid/far--infrared and mm wavelength range \citep{Dole06}.

Determining accurate dust masses for galaxies has been, traditionally, a difficult
step since measurements at infrared wavelengths shorter than $\sim$100--200~$\mu$m (as
accomplished by IRAS, ISO, and the Spitzer Space Telescope) are sensitive to the
inferred dust mean temperature(s) and the adopted dust emissivity. Wavelengths
longer than a few hundred $\mu$m probe the long wavelength tail of the modified 
Planck function for typical galaxy dust emission, and the inferred dust masses are, 
therefore, less sensitive to the dust temperature. Long--wavelength measurements 
also offer additional leverage for constraining, when used together with shorter 
wavelength infrared data, the dust temperature itself.

Recent applications of sub--mm data used to constrain the properties 
of the dust emission from galaxies include a nearby sample of about 100 galaxies 
observed at 450~$\mu$m and 850~$\mu$m with SCUBA \citep[The SCUBA Local Universe 
Galaxy Survey, a.k.a., SLUGS,][]{Dunne00, Dunne01, Seaquist04, Vlahakis05}, and an 
additional $\sim$20 galaxies from the SINGS sample \citep[the Spitzer Infrared 
Nearby Galaxies Survey,][]{Kennicutt03}, where Spitzer data in the range 
3--160~$\mu$m have been combined with SCUBA 850~$\mu$m detections \citep{Draine07}. 
More recently, sub--mm data have been combined with Spitzer data for 11 nearby 
galaxies from the SLUGS sample \citep{Willmer09}, thus increasing the total number 
of galaxies for which the dust masses are measured accurately from multi--wavelength 
spectral energy distributions, spanning from the mid--IR to the sub--mm. 
The SLUGS galaxies tend to have, on average, colder dust temperatures than the 
galaxies in the SINGS sample (of which NGC 1512/1510 is part), as estimated by 
\citet{Willmer09} on the basis of the far--infrared/sub--mm colors. 

The Spitzer data on nearby galaxies offer the unique opportunity to analyze 
the dust emission characteristics not only of whole galaxies, but also
within sub--galactic regions, thanks to its angular resolution in the range
2$^{\prime\prime}$--38$^{\prime\prime}$ for the 3--160~$\mu$m range. We
combine the SINGS data on the interacting galaxy pair NGC 1512/1510 with
data at 1.1~mm wavelength from ASTE/AzTEC, which has 28$^{\prime\prime}$
resolution, to investigate the infrared/mm spectral energy distribution (SED)
of dust for those two galaxies, and for regions within the larger NGC 1512.
Our goal is to derive accurate dust masses for the two galaxies, and to
compare the SEDs between sub-regions.

The interacting pair NGC 1512/1510 is of particular 
interest because its two member galaxies are remarkably distinct in nature: 
the primary galaxy, NGC 1512, is a large, metal-rich barred spiral, while 
its companion, NGC 1510, is a low metallicity blue compact dwarf (BCD) 
galaxy (Table~\ref{tab1}) which is expected to have less ability of self-shielding 
for the dust. Both galaxies are characterized by intense starburst activity in 
their central regions, and do not show evidence for presence of AGNs. 
This pair is thus ideal for investigations of dust SEDs, and their dependence 
on each galaxy's properties.

The paper is organized as follows: section 2 describes the data and, for
the ASTE/AzTEC data, provides some detail on the observations and data reduction; 
section 3
is devoted to the data analysis, including the photometry, and photometric   
corrections and uncertainties; section 4 presents the dust model fitting 
and results; the discussion and conclusions are in section 5.

\section{DATA}

\subsection{IRAC and MIPS Data} 

\begin{figure}
\figurenum{1}
\plotone{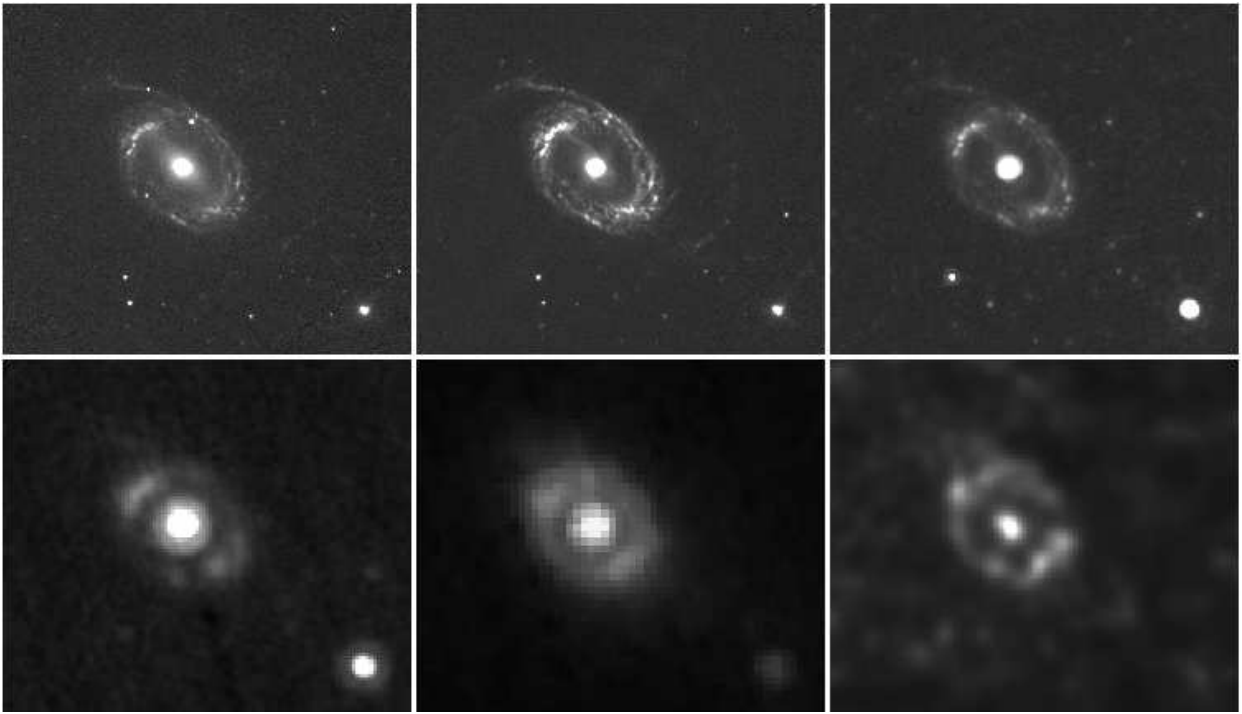}
\caption{Mosaic of the images of the galaxy pair NGC 1512/NGC 1510 
at 5.8, 8.0, 24, 70, 160~$\mu$m and 1.1~mm. North is up, East is 
left. The field of view of each panel is 8.7$^{\prime}\times$7.4$^{\prime}$.  
\label{fig1}}
\end{figure}

Spitzer maps with both the IRAC (3.6, 4.5, 5.8, and 8.0~$\mu$m) and MIPS (24, 
70, and 160~$\mu$m) are available for both galaxies, through the high level data 
products of the SINGS Legacy project \citep[Spitzer Infrared Nearby Galaxy 
Survey,][]{Kennicutt03}. Although the SINGS observations targeted NGC 1512 only, 
its companion galaxy is serendipitously present and detected in the same maps. 
The SINGS observation strategy, data reduction procedures, and map sensitivity 
limits are described in \citet{Kennicutt03} and \citet{Dale05}. The angular 
resolution is $\sim$2$^{\prime\prime}$ for the four \facility{IRAC} bands and 
$6^{\prime\prime}$, $17^{\prime\prime}$ and $38^{\prime\prime}$ for the 
\facility{MIPS} 24, 70 and 160 $\mu$m bands, respectively (Figure~\ref{fig1}).

\subsection{AzTEC/ASTE Data}

AzTEC is a 144 element bolometer array currently configured to observe
in the 1.1~mm atmospheric window~\citep{Wilson08a}.  Observations of
NGC 1512 were made with AzTEC on the Atacama Submillimeter Telescope
Experiment \citep[ASTE,][]{Ezawa04, Ezawa08} using a network observation 
system N-COSMOS3 \citep{Kamazaki05}, which provided a $28^{\prime\prime}$ 
circular main beam. During the period 7--11 October 2007 we performed 25 
lissajous observations identical to those described in \citet{Wilson08b} 
centered on ($\rm 04^h03^m54^s.28, -43^{\circ}20^{\prime}55.9^{\prime\prime}$). 
The coaddition of the 25 observations results in a roughly circular map with 
uniform coverage over the central 18$^{\prime}$ diameter with a depth of 
$\approx$1.5 mJy. The structure of NGC 1512 is well resolved, with 
a maximum signal-to-noise ratio $\sim$10 in the core, and $\sim$8 on the 
arms; its companion, NGC 1510 is detected with only marginal 
significance with a peak signal-to-noise of $\sim$3.5 (Figure~\ref{fig1}).

Atmospheric emission at 1.1~mm is the dominant noise source in our
AzTEC maps. Previous images made with the standard AzTEC pipeline
\citep{Scott08, Perera08} are optimized for point source recovery
and consequently do not preserve extended flux in the image.
We have developed our own iterative flux recovery technique that takes
advantage of the fact that the atmospheric noise is not correlated
between the 25 observations and that the atmosphere is transient in
time while the astronomical signal is stationary.  We begin by using a
variant of the principal component analysis (PCA) technique described
in \citet{Scott08} to remove atmosphere from the detector time streams.
Rather than the aggressive form of filtering described there 
--- which would result in a sharp spatial filter applied to 
this map --- we project just two eigenmodes from the 
detector--detector correlation matrix out of each 15 second span of 
time--ordered data.  This relatively weak filter is comparable in strength 
to a subtraction of the array average and gradient from each detector 
timestream sample. Once these initial ``cleaned'' data streams are produced 
we implement the following algorithm:

\begin{enumerate}

\item A coadded image of the field is made from the cleaned data streams.

\item The fluxes from pixels within 24$^{\prime\prime}$ of pixels with 
S/N$>$2.5 are preserved in the cleaned image while all other pixels are 
set to zero. We call this the ``current best sky'' image since it is the 
best estimate we have of a (noiseless) sky.

\item The ``current best sky'' image is recast into the time streams of 
the detectors and subtracted from the original detector time streams,
creating a set of residual time streams that, in principle, have the
same contamination from the atmosphere but less true astronomical
signal. We call these the ``residual observations''.

\item The ``residual observations'' are cleaned and mapped. The resulting
image is added to the ``current best sky'' image and a new ``current best
sky'' realization is produced using the same pixel flux criteria as
before.

\item This iterative process continues until no new pixels in the 
realization pass the pixel flux criteria.

\item At this point the severity of atmospheric filtering is increased.
That is, another detector-detector covariance matrix eigenvector is
projected out of the PCA basis for the atmospheric cleaning and the
iterations continue.

\item Iterations are complete when the full PCA technique for optimal
point source detection is realized (in this case cutting 7 eigenmodes
in the atmosphere cleaning step) and no new pixels pass the pixel flux
criteria \citep{Scott08}.

\end{enumerate}

\begin{figure}
\figurenum{2}
\plotone{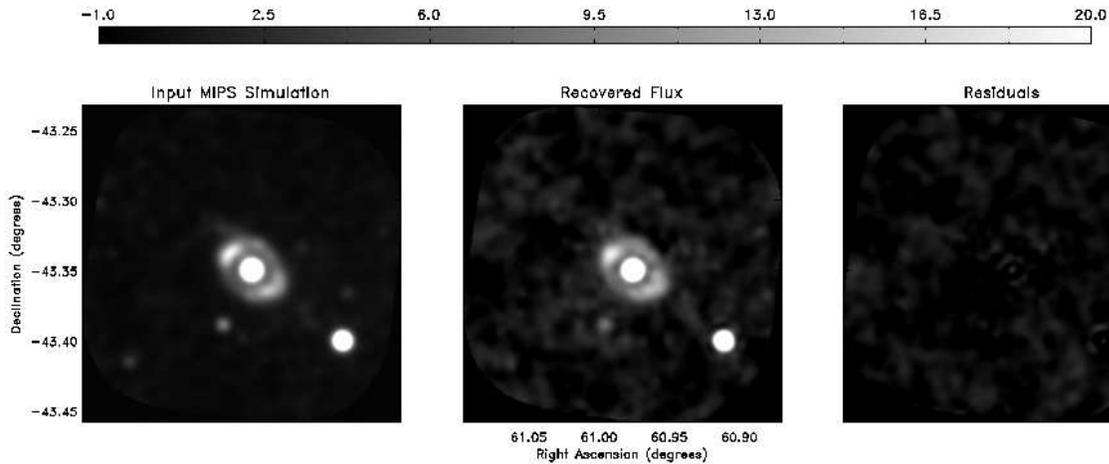}
\caption{(Left) Input MIPS 24~$\mu$m image scaled such that the extended flux
matches that measured in the AzTEC map.  This results in a peak flux
at the galaxy center of 133~mJy.  This image is injected into the
residual AzTEC timestreams as described in the text.  (Center) The
recovered image using the same FRUITLOOPS algorithm as used to produce
the AzTEC image of NGC 1512/1510.  (Right) The residual image (left image
minus center image). Negligible artifacts of 1-2 mJy from the AzTEC
analysis filters remain in the cores of the two galaxies due to the
bright initial image flux.  The faint extended flux is recovered to
well within the pixel noise of the map.
\label{fig2}}
\end{figure}

One benefit of this technique is that it results in a set of 25
residual timestreams that have been effectively ``cleared'' of
astronomical signal while leaving a good approximation of the true
atmospheric contamination behind. We take advantage of this by
injecting various forms of simulated signals into these residual
timestreams in order to perform tests on the effectiveness of the
iterative flux recovery technique. Figure~\ref{fig2} shows the results of
injecting a flux--scaled version\footnote{The flux of the MIPS map has
been scaled such that the extended features of the galaxy have a flux
that matches that of the AzTEC NGC 1512 map. This results in a peak
flux in the core of the galaxy which is significantly larger than what
we actually see at 1.1~mm, so this is a robust test of the
iterative flux recovery technique.} of the MIPS 24~$\mu$m map of NGC 1512,
smoothed with a 28$^{\prime\prime}$ Gaussian, into our residual timestreams 
and then repeating the identical analysis that was done for the AzTEC NGC 1512
observations. The residual map is consistent with the noise level in
the AzTEC NGC 1512 image and shows only minor residual systematics of
order 1--2 mJy at the locations of the cores of NGC 1510 and NGC 1512.
We perform the same aperture photometry analysis on the residual map
shown in Figure~\ref{fig2} as is done for the actual map (see Section 3.3) 
and find that our iterative flux recovery technique misses between 0.2\% 
and 12\% of the integrated flux depending on the scale-size of the
aperture. These values are small compared to the total integrated
flux and calibration uncertainty (17.1\% -- 128\%, see the next section).

\section{DATA ANALYSIS}

\subsection{Photometric Apertures}

\begin{figure}
\figurenum{3}
\plotone{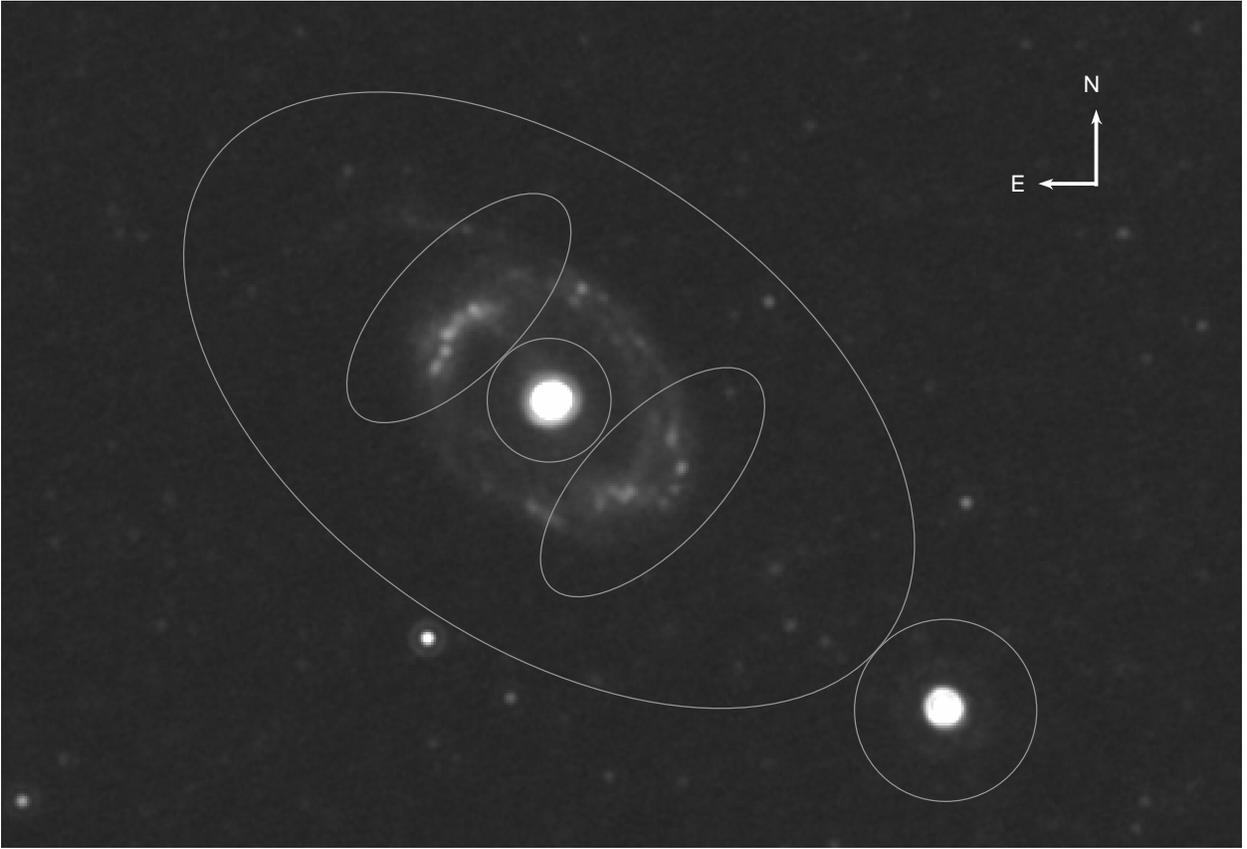}
\caption{The five photometric apertures used in this paper are shown as 
ellipses/circles, superimposed on the 24~$\mu$m map of the galaxy pair. 
The field of view is 11.4$^{\prime}$$\times$8.5$^{\prime}$.
\label{fig3}}
\end{figure}

In addition to measuring the fluxes from the whole galaxies, the infrared and 
mm maps have sufficient angular resolution to allow us to isolate substructures 
within the larger of the two galaxies, NGC 1512. We thus identify three separate 
subregions in this galaxy: the center, and two areas in the spiral arms 
(Figure~\ref{fig3}). For each of these regions, we define circular or ellipsoidal 
apertures, depending on the structure we are measuring. 

For NGC 1512 as a whole,  we define an elliptical aperture (Figure~\ref{fig3}), 
with major/minor axes $491^{\prime\prime}/287^{\prime\prime}$ and centered on the 
galaxy position defined in NED\footnote{The NASA/IPAC Extragalactic Database (NED) 
is operated by the Jet Propulsion Laboratory, California Institute of Technology, 
under contract with the National Aeronautics and Space Administration.}; the aperture's 
major and minor axes are matched to the optical major and minor axes and the position 
angle extracted from the KPNO/CTIO {\it BVRI} imaging campaign for the SINGS project 
\citep[Table~1 of][]{Dale07}. For NGC 1510, which is unresolved in the MIPS 
and mm images, and only marginally resolved in the IRAC bands, we define a single, 
circular aperture, with diameter 108.4$^{\prime\prime}$, centered again on the NED 
position of the galaxy. 

The size and location of the photometric apertures for the substructures in NGC 1512 
are more arbitrary in nature, and we define them based on the 1.1~mm map, taking also 
into account the lower resolution of the 160~$\mu$m image. The central region is defined 
by a circular aperture centered on the NED galaxy center, and with diameter 
73.6$^{\prime\prime}$, delimited by the size of the central emission region which is 
similar in the maps of the two longest wavelengths. The location and size of the two 
apertures along the spiral arms (Figure~\ref{fig3}) are chosen to encompass as much 
as possible of the northeastern and southwestern arm regions, while minimizing overlap 
with the central region. The arm apertures are thus elliptical in shape with major/minor 
axes 174$^{\prime\prime}$/78$^{\prime\prime}$. 

A constant background is determined and subtracted from each Spitzer image, using the 
peak value of the background pixels distribution \citep[see description in][]{Calzetti05}. The 
1--$\sigma$ uncertainty of the background band is, in unit of MJy sr$^{-1}$, 
[0.008, 0.012, 0.052, 0.053] for IRAC [3.6, 4.5, 5.8, 8.0]~$\mu$m, and [0.039, 0.36, 0.48] 
for MIPS [24, 70, 160]~$\mu$m, respectively. After background subtraction, photometric 
values are determined for each of the five regions in each band. 

\subsection{Photometric Corrections}

Our photometric measurements require a few corrections, due to the presence of: 
MIPS flux non--linearities, PSF wings outside the apertures, and, for IRAC, 
scattered light for extended sources.\footnote{http://spider.ipac.caltech.edu/staff/jarrett/irac/.} 
1.1--mm ASTE/AzTEC data require no aperture corrections.

MIPS 70~$\mu$m pixels with surface brightness above $\sim$66 MJy sr$^{-1}$ are subject 
to non--linearities that need to be corrected. The central pixel of NGC 1512 
is above this threshold, and we adopt the non--linearity correction formula published 
in \citep{Dale07} (equation 3), which is deduced from data given in \citep{Gordon07}. 
A correction factor of 1.011 is derived for the photometry of the whole galaxy, and 
1.023 for its central region. 

The extended source aperture corrections provided by the \facility{Spitzer}
Science Center for IRAC gives [0.911, 0.941, 0.798, 0.745] for NGC 1512, 
[0.928, 0.960, 0.901, 0.791] for its central region, [0.919,0.949, 0.866, 0.770] 
for the north arm, [0.919, 0.949, 0.866, 0.770] for the south arm and 
[0.920, 0.950, 0.869, 0.771] for NGC 1510 at [3.6, 4.5, 5.8, 8.0] $\mu$m 
respectively, with an average uncertainty $\sim$10\%.

We adopt the MIPS aperture corrections ([1.04, 1.05, 1.06] at [24, 70, 160] 
$\mu$m for NGC 1512 given in \citet{Dale07}. For the other apertures
we evaluate the corrections by measuring the wings of  MIPS PSFs for the 
relevant aperture sizes. The aperture correction for the central region of NGC 1512 
and for NGC 1510 are  [1.06, 1.13, 1.53]  and [1.04, 1.08, 1.34]  at  [24, 70, 160] 
$\mu$m, respectively. We derive aperture corrections of  [1.04, 1.08, 1.28] for 
the arm regions, which are roughly extrapolated from circular apertures inscribed 
and circumscribed to the minor and major axes, respectively, of the aperture ellipses.

\subsection{Uncertainties}

IRAC Calibration uncertainties are 5\%--10\% for 3.6 and 4.5~$\mu$m bands
and 10\%--15\% for 5.8 and 8.0~$\mu$m bands \citep{Dale07}. To be conservative, 
in this analysis we employ the upper limits of those ranges. For MIPS data, 
$4\%$, $7\%$ and $12\%$ are adopted at 24, 70 and 160~$\mu$m respectively
\citep[][]{Engelbracht07, Gordon07, Stansberry07}. The AzTEC/ASTE 1.1mm image 
calibration has an uncertainty of $\sim$8\%.

Aperture corrections are $\sim$10\% uncertain for IRAC channels. For the
MIPS bands, the aperture correction uncertainties are as small as a few percent, 
and our estimates of [$5\%$, $5\%$, $8\%$] at [24, 70, 160] $\mu$m are close to 
previous estimates \citep{Dale07}. 

The background noise levels in the \facility{Spitzer} images are found by performing a 
Gaussian fit of the pixel-value distributions. 
For the millimetric image, we test for false detections and fluctuations  by performing 
photometry of the background in the map using $\sim$8000 slightly 
displaced consecutive apertures (with the same area as each of the apertures used 
for source photometry) to cover the whole image, but excluding sources and the map edges. 
A Gaussian fit is then applied on the resultant histogram of
``background'' photometry and the dispersion of the background fluctuation is obtained
accordingly. This strategy results in relative 1-$\sigma$ total uncertainties at 1 mm 
(errors in absolute flux calibration and background estimation added in quadrature) of 
17.1\% (whole), 20.9\% (core), 24.8\% (northeast arm), 26.7\% (southwest arm) for 
NGC 1512, and 128\% for the entire NGC 1510. Total uncertainties 
computed in this fashion are listed in Table~\ref{tab2} for all the IRAC, MIPS and
AzTEC bands.

\section{MODEL FITTINGS AND RESULTS}

The silicate--graphite--PAH dust model of \citet{DraineLi07} is used 
to fit, via $\chi^2$ minimization, the flux densities of the two galaxies and of 
three sub-galactic regions in all eight bands. Parameters in the models are: 
the PAH mass fraction $q_{PAH}$, the lower cutoff of the starlight intensity 
distribution, and the fraction of the dust heated by starlight. The dust composition 
is assumed to be a mixture of carbonaceous grains and amorphous silicate grains, with 
size distributions consistent with the observed extinction curve in the local Milky Way 
\citep{Weingartner01}, along with varying amounts of PAH molecules. All charged PAHs 
(singly or multiply, positively or negatively) of a given size are assumed to have the 
same cross sections, but those of neutral and ionized PAHs are assumed to be different. 
This model employs a fixed shape for the PAH size distribution, a fixed PAH ionized 
fraction, and a fixed spectral shape of the illuminating starlight, which result in 
fixed ratios of PAH emission band strength. The stellar continuum contributing to the 
mid--infrared emission is approximated by a 5000~K blackbody \citep{DraineLi07}.

\begin{figure}
\figurenum{4}
%    \hspace{-.5mm}
    \includegraphics[totalheight=3cm,angle=0,origin=c,scale=1.4]{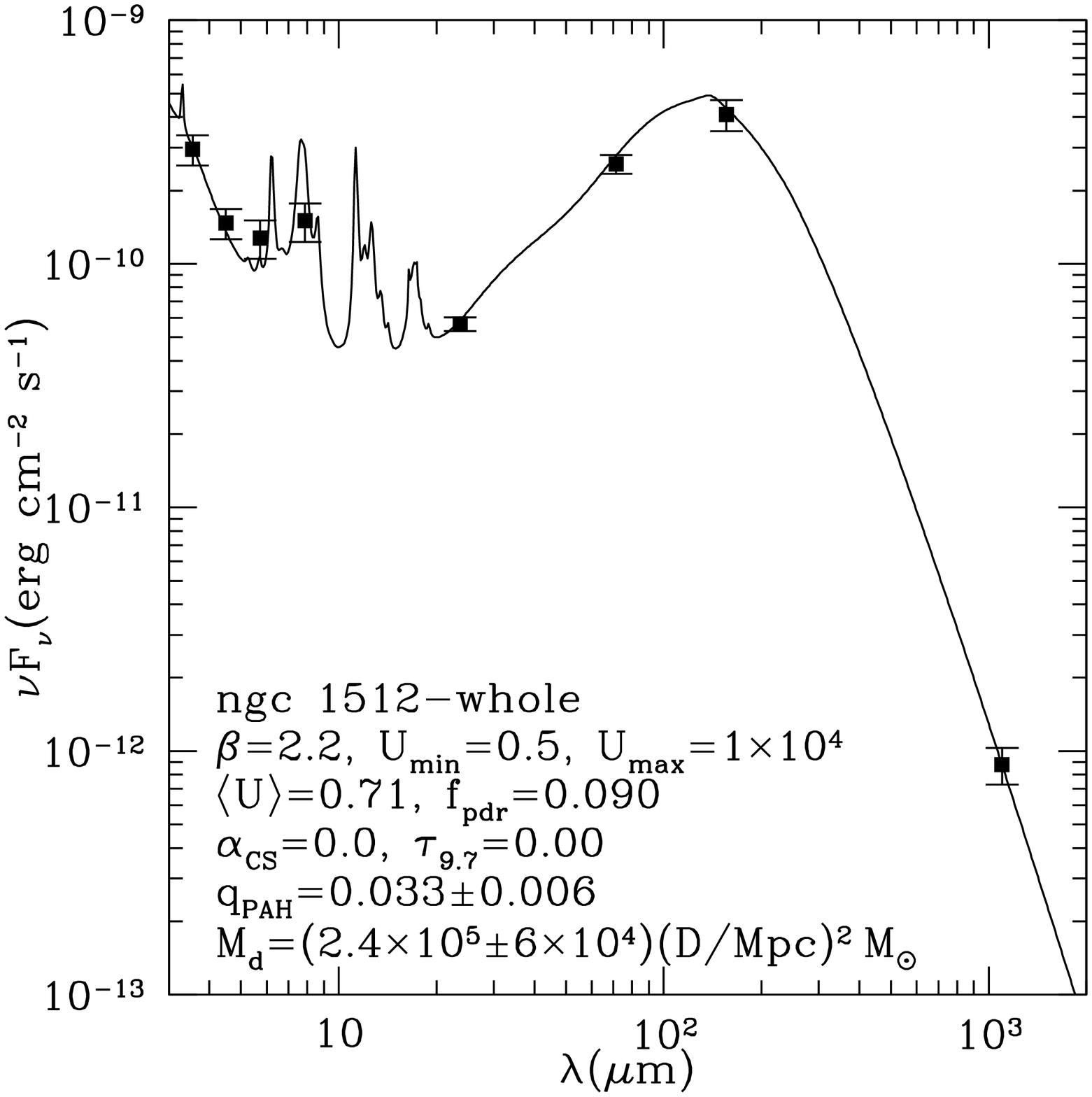}%
    \includegraphics[totalheight=3cm,angle=0,origin=c,scale=1.4]{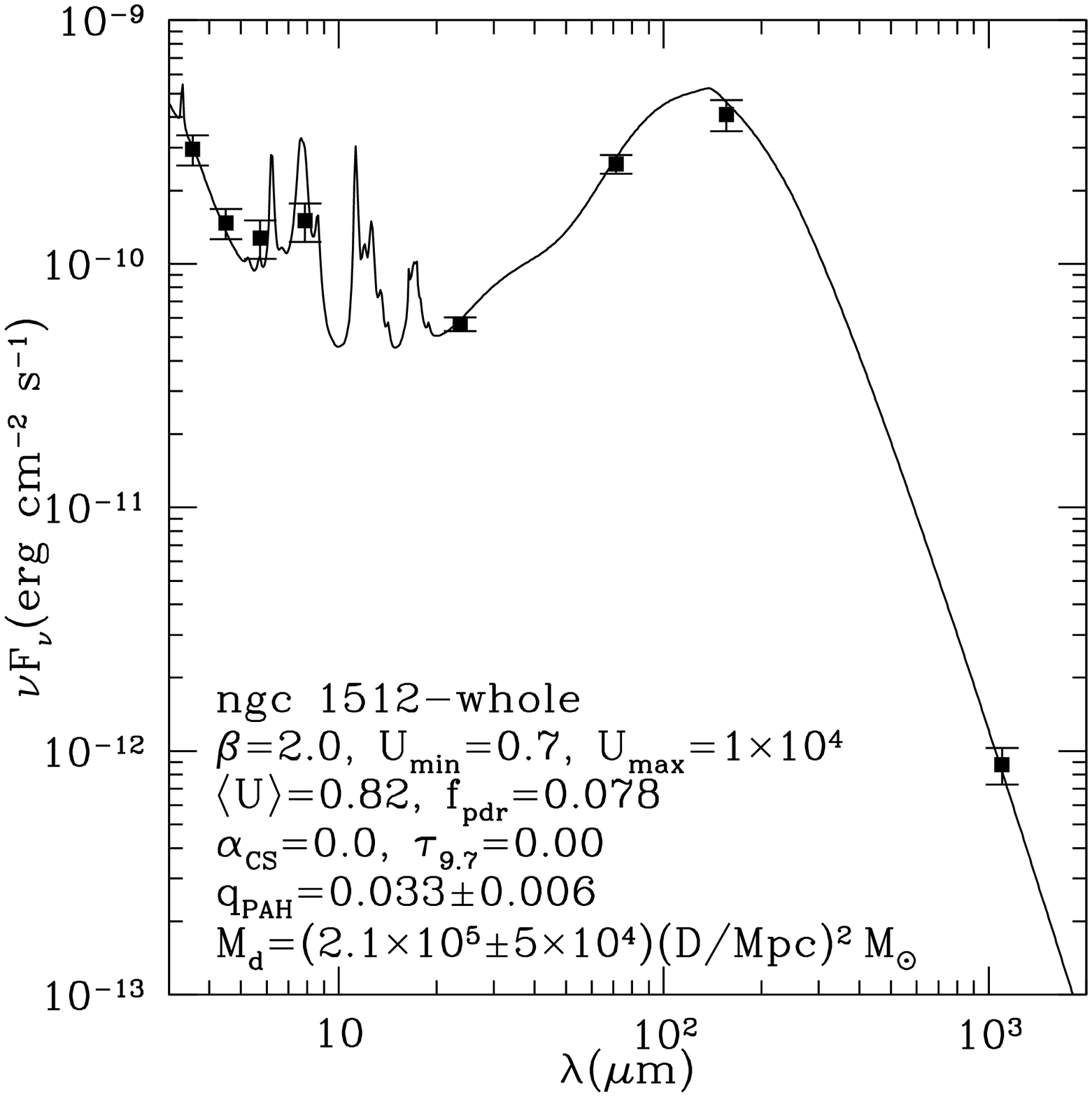}
    \hspace{4mm}
    \includegraphics[totalheight=3cm,angle=0,origin=c,scale=1.4]{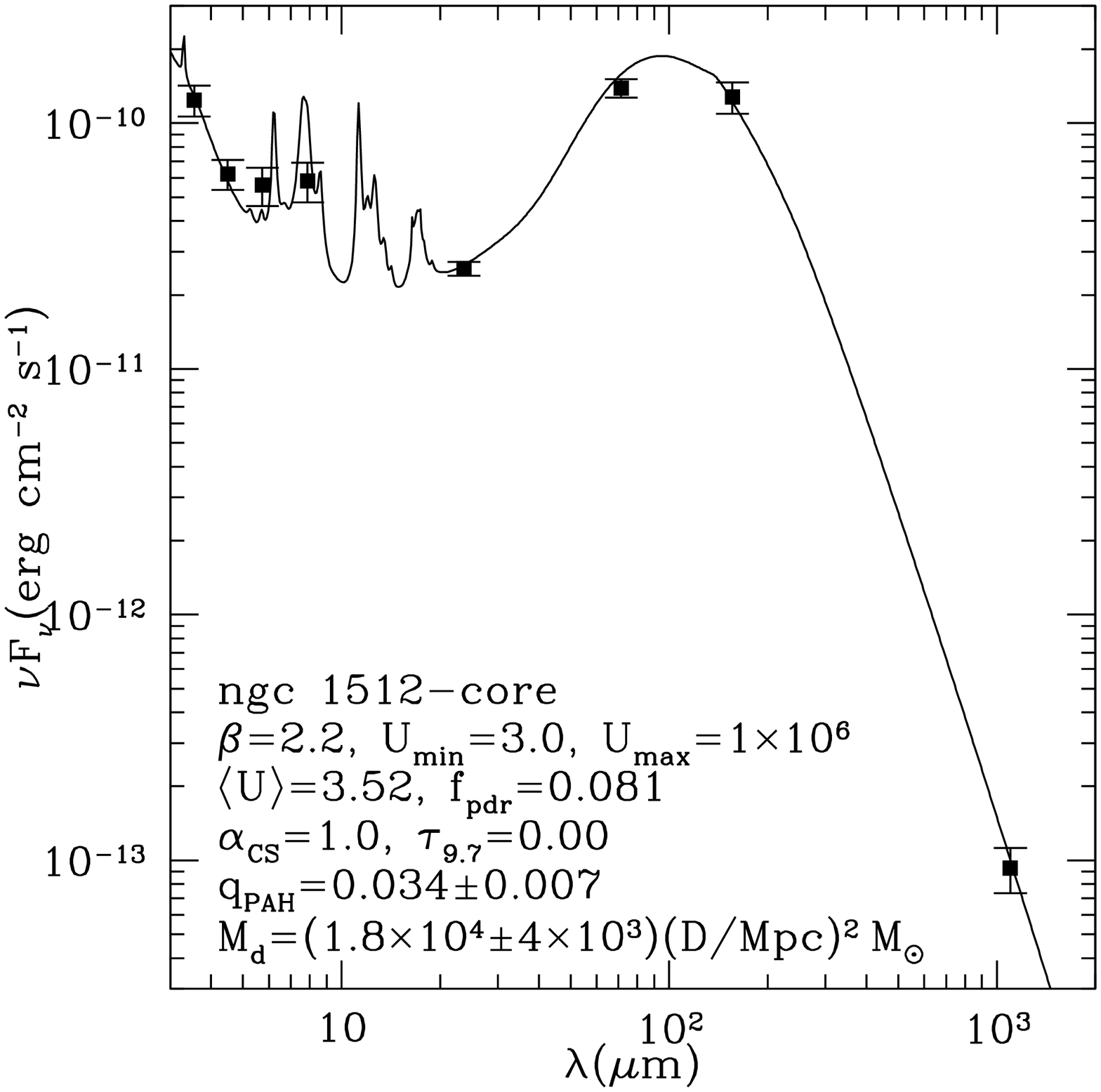}%
    \includegraphics[totalheight=3cm,angle=0,origin=c,scale=1.4]{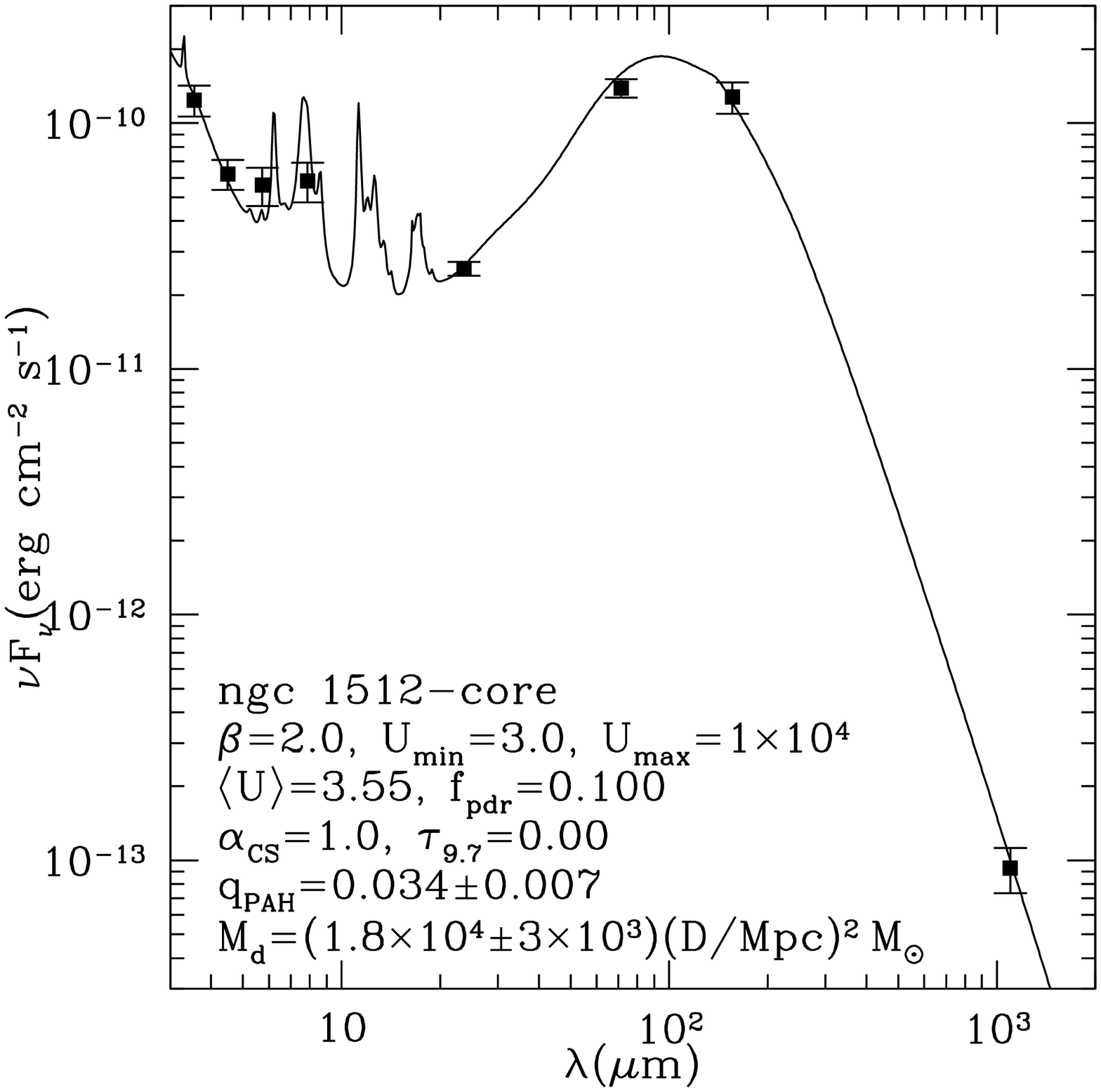}
    \hspace{4mm}
    \includegraphics[totalheight=3cm,angle=0,origin=c,scale=1.4]{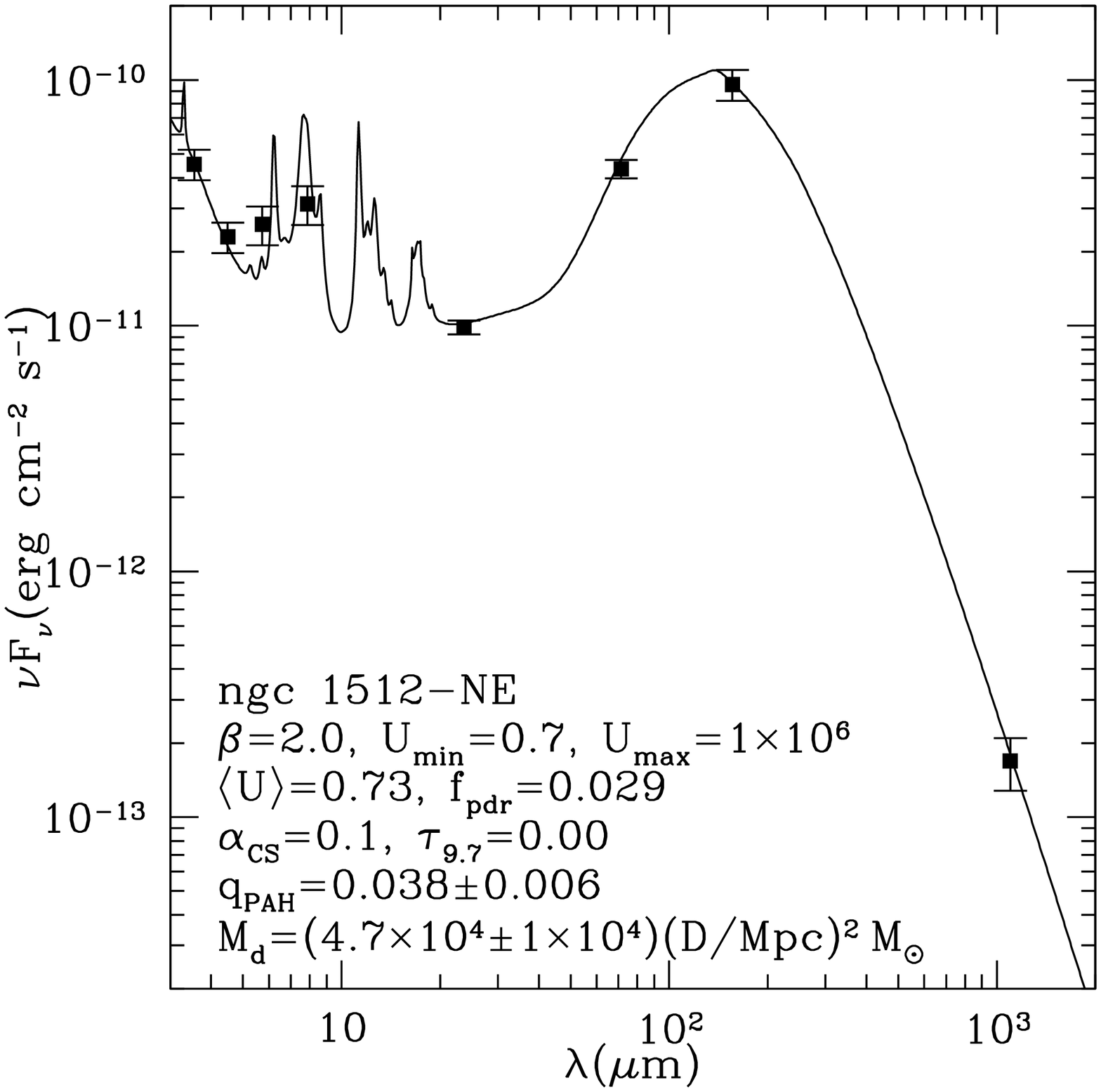}%
    \includegraphics[totalheight=3cm,angle=0,origin=c,scale=1.4]{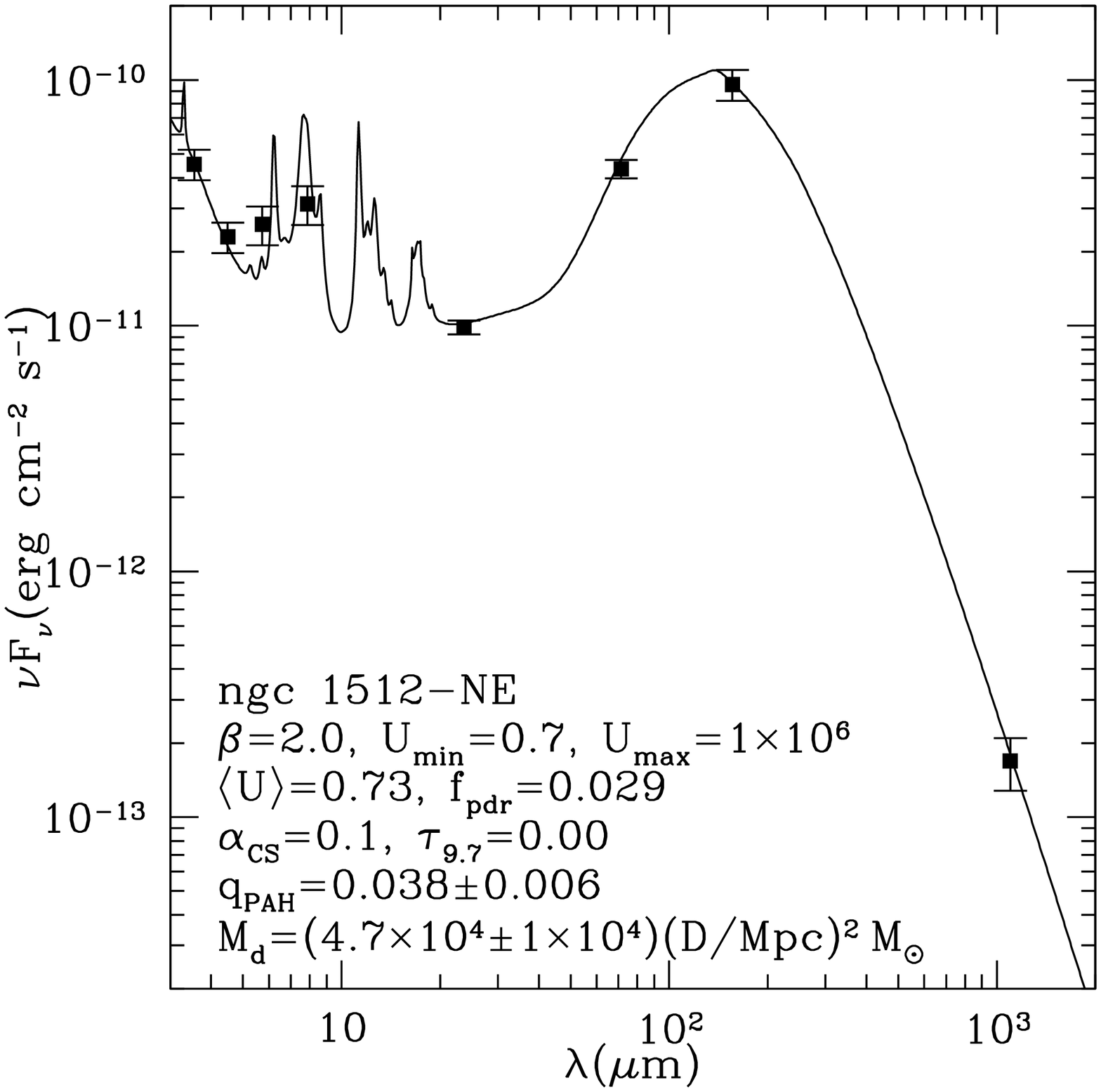}
    \hspace{4mm}
    \includegraphics[totalheight=3cm,angle=0,origin=c,scale=1.4]{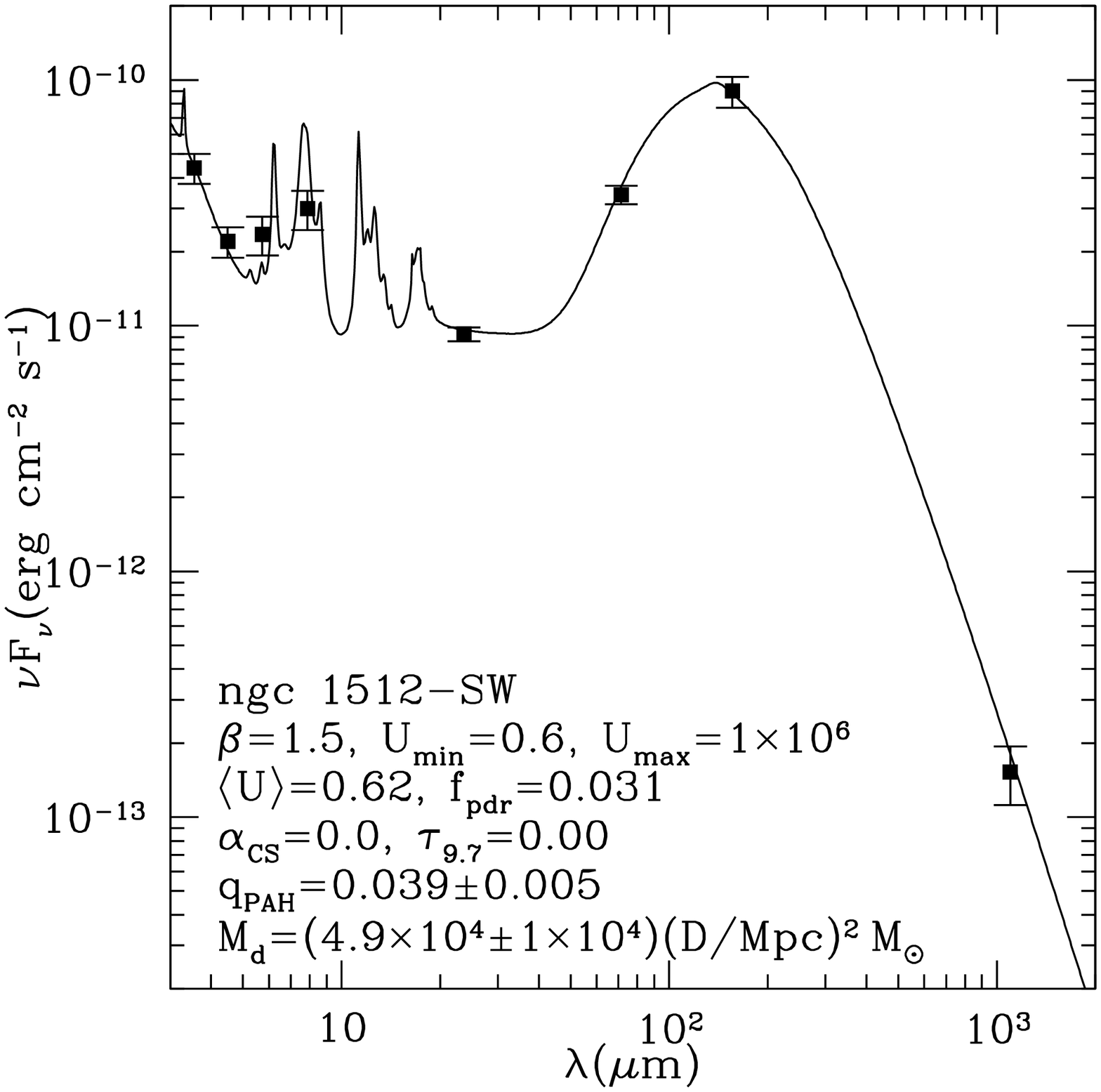}%
    \includegraphics[totalheight=3cm,angle=0,origin=c,scale=1.4]{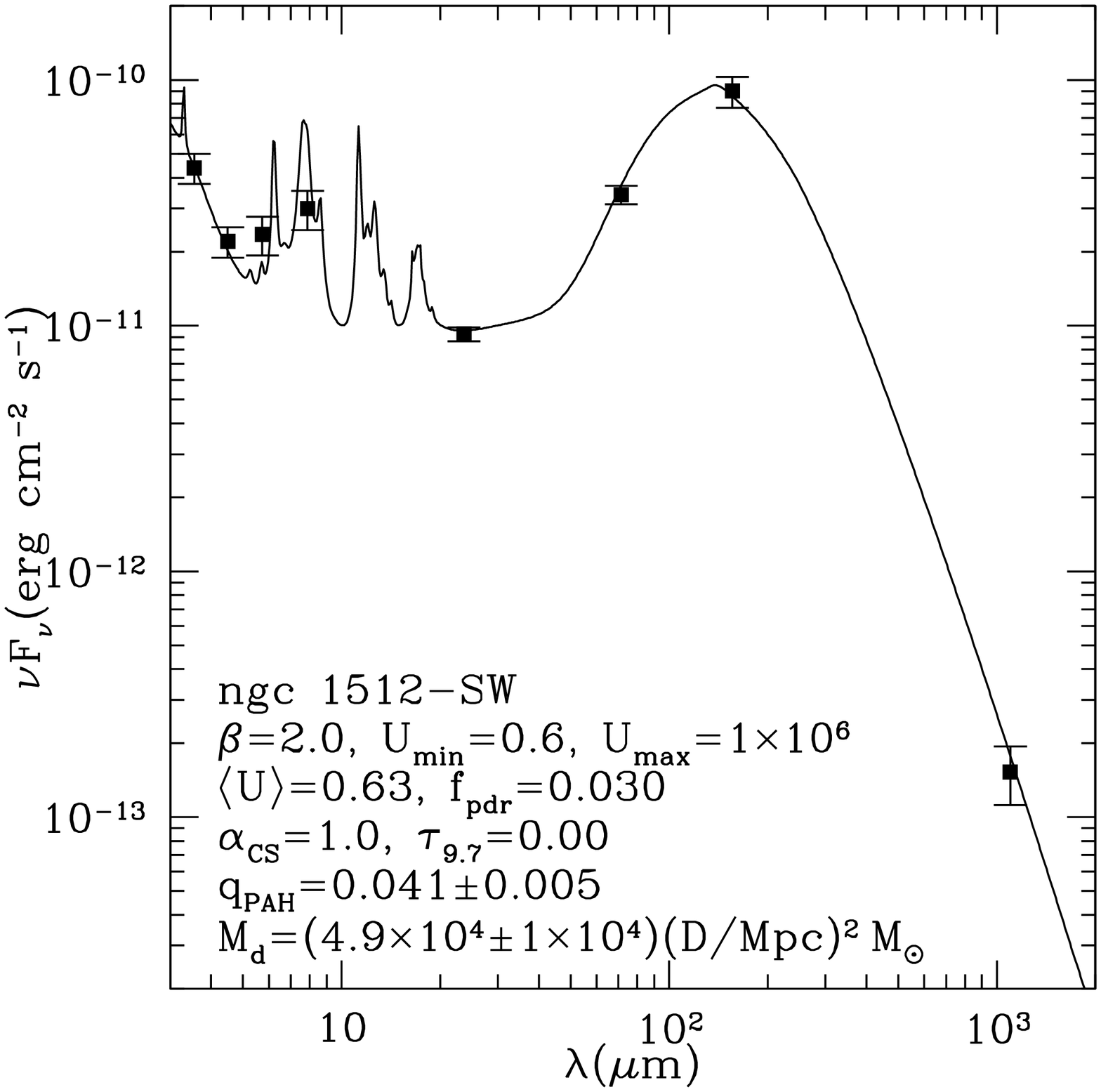}
    \hspace{4mm}
    \includegraphics[totalheight=3cm,angle=0,origin=c,scale=1.4]{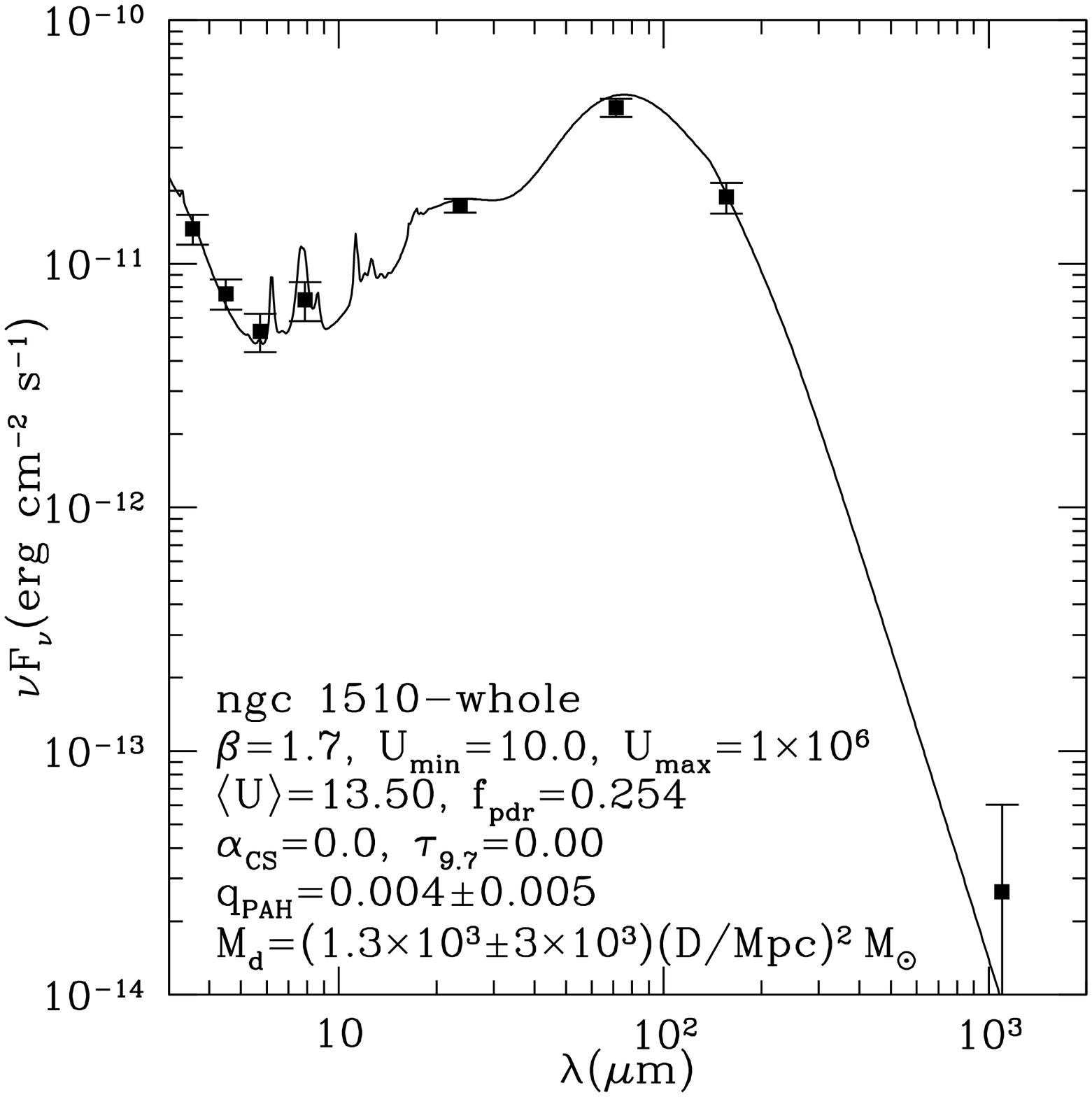}%
    \includegraphics[totalheight=3cm,angle=0,origin=c,scale=1.4]{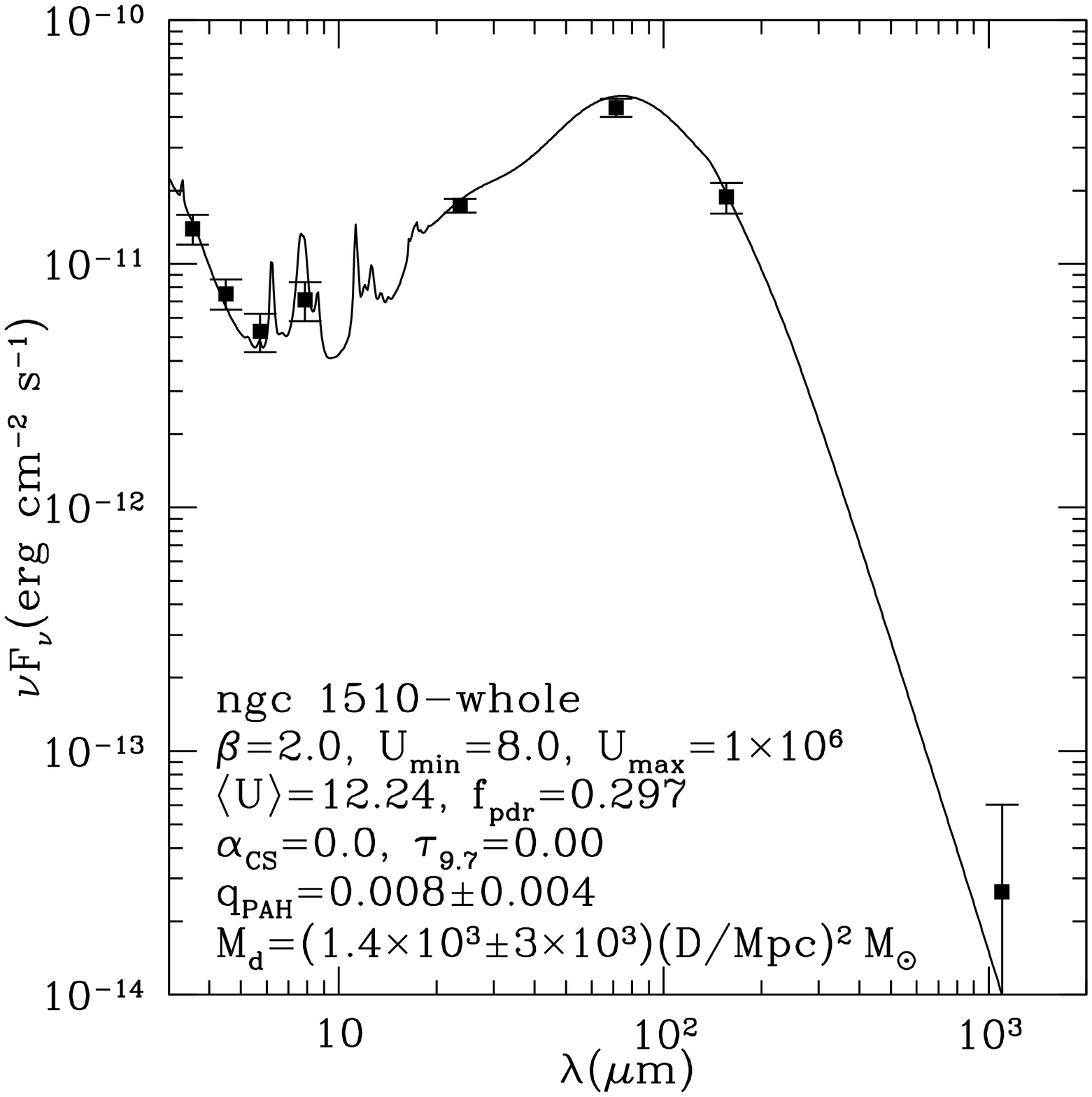}
    \caption{The SEDs of NGC 1512, its central region, its northwest and southwest 
arm regions, and its companion NGC 1510. All the SEDs are fitted by the model 
described in \citet{DraineLi07}. The results of two approaches of the fit are shown in 
a parallel manner: the left panel in each pair allows the parameter $\beta$ to float 
between 1.5 and 2.5 freely, but it is fixed for the right. However, the results sometimes 
coincide with each other.}
\label{fig4}
\end{figure}

The comparison between data and best-fit models are shown in Figure~\ref{fig4}, together 
with the resulting model parameters. The SEDs of NGC 1512 and its regions  are well 
represented by the models, yielding dust masses $\rm M_d$$\sim$2.4$\times$10$^7$~M$_{\odot}$ 
for the whole galaxy, and in the range $\sim$2--6$\times$10$^6$~M$_{\odot}$ for the 
regions within the galaxy (see Table~3), using the distance listed in Table~1. 
NGC 1510 is less well fit by the model, mainly owing to the combination of a high 
70/160~$\mu$m ratio (which implies a high effective dust temperature) together with a 
highly uncertain 1.1~mm flux (whose value  would imply presence of large amounts of 
cooler dust, see below). 
%We ascribe the somewhat discrepant measurements to  the immature AzTEC pipeline for extended source processing. 
Despite the uncertainties, the models still provide a reasonable fit to the NGC 1510 SED, 
with a resulting dust mass $\rm M_d$$\sim$1.7$\times$10$^5$~M$_{\odot}$, or about 150 times 
less dust mass than its companion galaxy. This difference is reduced to a factor of 45, 
if we use the upper bound of the large uncertainty in NGC 1510's dust mass determination. 

%This value is about 3.7 times lower than what is expected from the simple scaling between 
%the two galaxies' oxygen abundances (Table~1), but the uncertainty on the dust mass is at least 
%a factor of 2 (Figure~3); taking a 2$\times$ higher dust mass reduces the discrepancy relative 
%to the scaling of the abundances to less than a factor of 2. 

%{\bf BRUCE: would you have more to add?}

As a test, we have also fit the far--infrared (70 and 160~$\mu$m) and mm data to a very 
simple prescription of two modified blackbodies, with temperatures and relative 
intensities derived through $\chi^2$ minimization (Table~4). The power-law index of 
the dust emissivity is fixed to the value $\epsilon$=2, following the conclusions of 
both \citet{Dunne01} and \citet{Willmer09} who determine that this value produces better 
fits to the SEDs of the galaxies they consider than lower power-law index values. 
Overall, this is a very crude prescription, with 4 free parameters (warm dust temperature, 
$\rm T_w$, cool dust temperature, $\rm T_c$, the relative contribution of the two dust components 
to the observed infrared--mm SED, and the total luminosity), thus the fit is not constrained 
by our three data points (70~$\mu$m, 160~$\mu$m and 1.1~mm). Indeed, we only derive these 
quantities for comparison purposes. However, in order to guide our physical intuition, we 
use the masses derived through the physical models of \citet{DraineLi07} as our ``target 
values'' for the simple two--temperature prescription. For both galaxies and all regions, 
we find that the colder of the two temperatures is in the range $\rm T_c$$\sim$14--16~K 
(Table~4). As a comparison, within the context of \citet{DraineLi07} model, seeing that 
stronger radiation fields result in hotter dust grains, the lowest dust temperature can 
be inferred from the lower cutoff of the starlight intensity distribution $U_{min}$. The 
best--fit model of the whole NGC 1512 is parameterized by $U_{min}$=0.7, corresponding to 
a lowest dust temperature $\sim$16~K, which is consistent with our result of 
two--temperature fittings as mentioned above. While for NGC 1510, $U_{min}$=8.0 and therefore 
a lower temperature limit is found to be 24~K, which is somewhat higher than that of 
the cooler dust component ($\sim$15~K) obtained with the two--temperature fitting 
strategy. However, we do not regard this as a discrepancy, given that the 1.1~mm data point of NGC 1510 is highly uncertain and  that both temperature 
values are still consistent with the fact that active star--forming galaxies 
appear to have cool dust components with temperatures of--order 20~K (see the Discussion
section).
For the warmer dust component, we have a markedly different behavior for the two 
galaxies: NGC 1512 and its regions are described by ``warm'' dust temperatures in the range 
$\rm T_w$$\sim$20--25~K, with the higher value being associated with the central region of the 
galaxy; NGC 1510 requires a warmer temperature than NGC 1512, $\rm T_w$$\sim$36~K, to fit its 
SED. This is a well--known result:  low--mass galaxies tend to have higher effective dust temperatures than large spirals 
\citep{Hunter89,Dale05}. The total dust masses derived with the two--temperature method 
are higher than those derived from the more accurate fits using the \citet{DraineLi07} models, 
by a factor 1.6--2.5 in NGC 1512, and by a factor $\sim$9 in NGC 1510, or 2.8 if the upper 
bound of its dust mass is applied (Table~4). Assumptions of lower dust 
emissivity would yield even higher dust masses from the two-temperature method
than those derived with the \citet{DraineLi07} method (see the Conclusions).
% {\bf GUILIN: can you please check this?}
  
For NGC 1512 as a whole the warmer dust component includes $\sim$7\% of the total 
dust mass in the galaxy, and the dominant contributor by mass is the cooler dust component 
(Table~4). However, in the sub--galactic regions of NGC 1512, the warm dust represents a larger 
fraction of the total dust mass, between $\sim$17\% and $\sim$40\%. This is consistent with the fact 
that the sub--galactic regions have been selected to encompass actively star forming areas in the 
galaxy, and are likely biased towards higher warm dust contents than the average across NGC 1512.  
In NGC 1510 the mass--dominant dust component is the cool one, with a mass that is about 
36 times larger than that of the warmer component; this is perhaps not surprising as the 
higher warm dust temperature for this galaxy relative to NGC 1512 naturally leads to a much higher 
emissivity, corresponding to much smaller warm dust mass. Finally, we remark again that the 
results for this galaxy are highly uncertain, due to the low--significance of the mm flux determination. 

  %There is still a discrepancy, by about a factor 
  %of 2, between the ratio of the dust masses of the two galaxies and the ratio of their oxygen 
  %abundances. 

\section{DISCUSSION}

The availability of mm observations for the nearby, star--forming galaxy pair NGC 1512/1510 
indicates that the observed infrared/mm SED of both galaxies and of the three 
sub--galactic regions in NGC 1512 analyzed in this paper are consistent with the presence of 
a relatively cool dust component ($\rm T_c$$\sim$14--16~K, Figure~\ref{fig3}), if a power-law
index of the dust emissivity $\epsilon$=2 is assumed. Presence of a non--negligible cool 
dust component with temperature below 20~K has already been suggested for nearby elliptical 
galaxies \citep[][]{Haas98, Klaas01, Leeuw04}. In M31, a cold dust component with a 
temperature of 16$\pm$2~K seems to be present as well \citep[][]{Odenwald98, Genzel00}. 
Confirming the actual existence of this component in our case is difficult, due to the 
crudeness of the two--temperature models and to the sparseness of the SED sampling. In general,
data between 160~$\mu$m and $\sim$1~mm are not available for galaxies, although a few galaxies 
(including NGC 1512) have been recently observed at wavelengths between 250~$\mu$m and 500~$\mu$m with 
BLAST \citep{Wiebe09}, and more will become available thanks to the observations with the Herschel Space 
Telescope. Interestingly, even very actively star forming galaxies, like LIRGs and ULIRGs, do seem to 
require some cool dust component, with temperature of--order 20~K, about a factor 2--2.5 
smaller than the temperature of the warm dust component that dominates the total 
luminosity \citep{Dunne01}.

The observed far--infrared/mm SEDs of the two galaxies cannot be easily 
fit by a single modified blackbody, without requiring that the dust  has a power--law 
index $<$2: NGC 1512 requires $\epsilon$$\sim$1.3 and NGC 1510 $\epsilon$$\sim$1.1, the latter 
value being at the margin of the acceptable range for dust emissivity (1$\le$$\epsilon$$\le$2, 
Seki \& Yamamoto 1980). Less stringent are the fits of the sub--galactic regions, where the 
single--temperature assumption requires $\epsilon$$\sim$1.5--1.8, marginally consistent with 
$\epsilon$=2; indeed, these regions have been selected to be actively star--forming, so it is 
not surprising that their infrared emission is dominated by the warm dust component. 

Dust emissivity values $\epsilon$$<$2 have been derived for dust emitting regions in the 
Magellanic Clouds \citep{Aguirre03}, assuming a single temperature for the modified blackbody; 
however, two--temperature models with $\epsilon$=2 appear to produce equally good fits to the 
data \citep{Aguirre03}. Furthermore, the multi--wavelength infrared/sub--mm data of  ULIRGs 
and SLUGS galaxies seem to indicate that better fits are obtained when multiple--temperature 
modified blackbody emission with emissivity having a power-law index $\epsilon$=2 
\citep{Dunne01,Willmer09}.  The presence of multiple temperature dust components is supported 
by the recent conclusions of \citet{Rieke09} and \citet{Calzetti09}, who find evidence for 
dust self-absorption in the mid--infrared SEDs of LIRGs, which then requires re-emission at 
longer wavelengths. These results provide ground for an ``onion--peel'' scenario, where 
different dust layers are heated at different temperatures. 

Single--temperature fits yield values around 20--25~K and $\sim$36~K
for the warm dust components of NGC 1512 (and its sub--regions) and NGC 1510, respectively. This 
is consistent with NGC 1510 being, on average, more actively star--forming than NGC 1512, when 
considering the star formation rate density (Table~1), and also being less metal--rich, which 
leads to less ability of self--shielding for the dust \citep{Calzetti00}. We also notice
a higher temperature of the warmer dust in the center of NGC 1512 (25~K) compared to that of
the arm regions of the same galaxy (20--21~K), implying a stronger radiation field in the 
galaxy center where more UV photons are reprocessed by dust grains.

The combination of a long wavelength baseline for the dust emission (thanks to the mm 
data) and the physically--motivated models \citep{DraineLi07} enables us to derive accurate 
dust masses for NGC 1512 and its subregions: typical 1~$\sigma$ uncertainties are in the 
range 19\%--25\% (Table~3). Less secure is the dust mass of the fainter NGC 1510, with 
a factor $\sim$3 uncertainty, due to its faint mm emission.  We should stress that our quoted 
accuracy is within the context of the \citet{DraineLi07} models. Even with this uncertainty, 
we are in a position of deriving dust--to--gas ratios for the galaxy pair.

The two interacting galaxies, located in the southern sky, do not have any CO (tracer of 
molecular gas) observations; therefore we will be limited to the derivation of $\rm M_d/M_{HI}$
ratio. Interferometric VLA observations exist for the pair \citep{Thornley06}, in addition 
to single--dish data from the Parkes Telescope 
\citep[as part of the HIPASS survey,][]{Koribalski04}. We use the latter data with low angular 
resolution to recover the diffuse emission in the galaxy pair which is missed by 
the interferometric data. The Parkes data contain a total emission of 259$\pm$17 Jy km/s  
at 21~cm, and by comparing this figure with the total flux ($>$3$\sigma$) contained in 
the VLA map, we estimate that the interferometric observations miss $\sim$52\% of 
the total {\sc H i} gas mass. We assume this ``diffuse'' component to be homogeneously distributed 
within the area occupied by the two galaxies; this assumption is very simplistic, but we 
expect that its impact on the individual regions (which are dominated by the clustered {\sc H i} 
detected by the VLA) will not be large, except for the center of NGC 1512 where the {\sc H i} flux
is weak. We then measure the 21~cm emission within each of the five regions defined in Table~2 from 
the VLA image, include the correction for the ``diffuse'' gas derived above, and derive {\sc H i} 
masses using the conventional formula 
$\rm M_{HI}/M_{\odot}$= 2.36$\times$10$^5$~$\rm I_{21cm}D^2_{Mpc}$, where the integrated intensity 
$\rm I_{21cm}$ is in Jy km/s and the distance in Mpc. 

When enclosing the entire {\sc H i} emission from the NGC 1512/NGC 1510 pair,  we 
obtain $\rm M_d/M_{HI}$=0.0034$\pm$0.0008, consistent with the 
value 0.0028 found by \citet{Draine07}. If only the {\sc H i} emission inside our photometry aperture 
for NGC 1512 itself (see description in the data analysis section) is taken into account, the 
$\rm M_d/M_{HI}$ ratio becomes an order of magnitude higher (0.037). Significantly higher values 
are obtained for the northern ($\rm M_d/M_{HI}$=0.089) and southern ($\rm M_d/M_{HI}$=0.079) arm 
regions, and for the core ($\rm M_d/M_{HI}$=0.13). These values should be, however, considered upper 
limits to the actual dust--to--gas ratios, since we do not have $\rm H_2$ measurements, and molecular gas is 
likely to be an important component in these three actively star--forming regions. Based 
on spatially--resolved CO studies of other massive star--forming galaxies 
\citep[e.g.,][]{Kennicutt07, Leroy08}, we estimate the $\rm H_2$ contribution to the 
gas content of the sub--galactic regions in NGC 1512 to 
be at least of the same order as the {\sc H i} itself. Using the expected proportionality between the 
dust--to--gas ratio and metallicity in galaxies, 
$\rm M_d/M_{gas}$$\approx$0.010(O/H)/(O/H)$_{\rm MW}$, from \citet{Draine07}, we can tentatively 
compare this formula with our derived dust and {\sc H i} masses. Using the metallicity 
value in Table~1, we find for NGC 1512, $\rm M_d/M_{gas}$=0.011, which is lower than the 
$\rm M_d/M_{HI}$ value we determine  the galaxy. We attribute this discrepancy to the 
absence of molecular gas data. For NGC 1510, we derive 
$\rm M_d/M_{HI}$=0.0027, and, from its metallicity (Table~1) and \citet{Draine07}'s formula, 
an expected $\rm M_d/M_{gas}$=0.0042. Again, although the two numbers differ by about 
a factor of 2, lack of molecular gas information hampers detailed comparisons. 

While $\rm M_d/M_{HI}$ ratios for individual regions and/or the whole galaxies appear 
consistent or larger than what one would expect for $\rm M_d/M_{gas}$ as given by 
\citet{Draine07}, the dust--to--gas ratio of the galaxy pair is lower, by about a factor of 3, 
than what expected, based on the same formulae.  This `deficiency'  would be exacerbated by 
the presence of significant molecular gas in the system. We speculate this deficiency to be 
due to the presence of a large fraction of {\sc H i} not associated with the star--forming disk(s) 
and/or regions within the two galaxies. A solution to this issue will require both higher sensitivity 
{\sc H i} maps and the availability of CO maps to measure the contribution and distribution of 
the molecular gas. 

\section{SUMMARY}

The combination of Spitzer mid/far--infrared images with ground--based ASTE/AzTEC 1.1~mm data
for the galaxy pair NGC 1512/1510 has enabled us to derive accurate ($\sim$19\%--25\% uncertainty) 
dust masses for the large spiral NGC 1512 and three of its subregions (the center and the two 
spiral arms), and constrain the dust mass  of the low--metallicity dwarf galaxy NGC 1510 to within 
a factor of 3. The total dust mass of NGC 1512 is 2.4$\times$10$^7$~M$_{\odot}$ and it is about 
150 times smaller in NGC 1510, in agreement with the former being a large spiral galaxy and the 
latter a small compact dwarf galaxy. 

In both galaxies, when the SEDs are fitted with simplistic two-component modified Planck 
functions, the majority of the dust mass is found to have a relatively cool temperature, 
$\sim$14--16~K, similar to what found for the nearby galaxy M31. Conversely, in the central 
region of NGC 1512 the cool and warm ($\rm T_w$$\sim$25~K) dust contribute close to equal mass, 
reflecting the larger specific SFR of the starbursting center relative to the galaxy as a whole. 
The two arm regions display properties that are intermediate between the central region and the 
whole galaxy in terms of ratio between the warm and cool dust mass, with values in the range 
0.2--0.3. 
 
The dust--to--gas ratio in NGC 1512 (estimated using only {\sc H i}, because of lack of CO 
emission data for the pair) is about 3 times below the expectation for a galaxy with a 
metallicity similar to the Milky Way, as already remarked in \citet{Draine07}. The addition 
of $\rm H_2$ to the gas census would only make the discrepancy worse. Conversely, the subregions 
(center and arms) in the galaxy show high $\rm M_d/M_{HI}$ ratios, $\sim$0.08--0.09 for the 
arms and $\sim$0.13 in the center, thus about 20 to 30 times larger than the galaxy as a 
whole, and consistent with the lack of $\rm H_2$ information. We speculate that much of the 
{\sc H i} included in the $\rm M_d/M_{HI}$ ratio estimate of the galaxy as a whole is not 
related to the star--forming disk.

\acknowledgments

The ASTE project is driven by the Nobeyama Radio Observatory (NRO),
a branch of National Astronomical Observatory of Japan (NAOJ),
in collaboration with the University of Chile, and Japanese institutes
including University of Tokyo, Nagoya University, Osaka Prefecture 
University, Ibaraki University, and Hokkaido University. A part of 
this study was supported by the MEXT Grant-in-Aid for Specially 
promoted Research (No.~20001003). KS was supported in part 
through the NASA GSFC Cooperative Agreement NNG04G155A.

\clearpage

\begin{table}
\begin{center}
\caption{Characteristics of the Galaxies.  \label{tab1}}
\begin{tabular}{lll}
\tableline\tableline
 & NGC 1512 & NGC 1510 \\
\tableline
Morphology\tablenotemark{a} &SB(r)ab &SA0$^0$ pec?; {\sc H ii} BCDG \\
$v_{\rm H}$ (km/s) &898$\pm$3 &913$\pm$10 \\
D (Mpc)\tablenotemark{b} &10.8$\pm$0.8 &11.0$\pm$0.8 \\
12$+$log(O/H)\tablenotemark{c} &8.71  & 8.31\\
$\Sigma_{\rm SFR}$\tablenotemark{d} & 0.001& 0.06\\
\tableline
\end{tabular}
%% Any table notes must follow the \end{tabular} command.
\tablenotetext{a}{Galaxy morphology and heliocentric
velocity are from the NASA/IPAC Extragalactic Database (NED).}
\tablenotetext{b}{Adopted distances.}
\tablenotetext{c}{Oxygen abundances, from \citet{Moustakas09} for NGC 1512 and 
from \citet{Storchi94} for NGC 1510.}
\tablenotetext{d}{The star formation rate density, in units of 
M$_{\odot}$~yr$^{-1}$~kpc$^{-2}$, from a combination of H$\alpha$ and 24~$\mu$m measurements for 
NGC 1512, to account for both the dust obscured and unobscured star formation 
\citep{Kennicutt09}, and from H$\alpha$ measurements for NGC 1510 \citep{Calzetti09}.}
\end{center}
\end{table}

\clearpage

\begin{table}
\begin{center}
\caption{Photometric measurements of the galaxy pair NGC 1512/1510. \label{tab2}}

\begin{tabular}{llllll}
\tableline\tableline
$\lambda(\mu m)$ & NGC 1512 & NGC 1512$_{\rm C}$\tablenotemark{\dagger} 
& NGC 1512$_{\rm NE}$\tablenotemark{\dagger} 
& NGC 1512$_{\rm SW} $\tablenotemark{\dagger} & NGC 1510 \\
\tableline
3.6~(mJy) &349$\pm$49   &147$\pm$21   &54$\pm$8      &52$\pm$7       &17$\pm$2 \\
4.5~(mJy) &220$\pm$31    &93$\pm$13   &34$\pm$5      &33$\pm$5       &11$\pm$2 \\
5.7~(mJy) &244$\pm$44   &107$\pm$19   &49$\pm$9      &45$\pm$8       &10$\pm$2 \\
7.9~(mJy) &394$\pm$71   &153$\pm$28   &82$\pm$15     &79$\pm$14      &19$\pm$3 \\
23.7~(mJy)&447$\pm$29   &202$\pm$13   &78$\pm$5      &73$\pm$5      &136$\pm$9 \\
71.4~(Jy)&6.24$\pm$0.54 &3.24$\pm$0.28 &1.03$\pm$0.09 &0.81$\pm$0.07  &1.04$\pm$0.09 \\
156~(Jy) &21.3$\pm$3.1 &6.65$\pm$0.96 &4.99$\pm$0.72 &4.68$\pm$0.68  &0.98$\pm$0.14 \\
1100~(mJy)&322$\pm$55    &34$\pm$7    &62$\pm$15     &56$\pm$15      &10$\pm$12 \\
\tableline
\end{tabular}
%% Any table notes must follow the \end{tabular} command.
\tablenotetext{\dagger}{NGC 1512$_{\rm C}$ represents the central region 
of NGC 1512, while the northeastern/southwestern arms are denoted
by NGC 1512$_{\rm NE}$/NGC 1512$_{\rm SW}$.}
%\tablecomments{}
\end{center}
\end{table}

\clearpage

\begin{table}
\begin{center}
\caption{Derived Quantities for NGC 1512 and NGC 1510 \label{tab3}}

\begin{tabular}{llllll}
\tableline\tableline
$Parameter$ & NGC 1512 & NGC 1512$_{\rm C}$\tablenotemark{\dagger} 
& NGC 1512$_{\rm NE}$\tablenotemark{\dagger} 
&\ NGC 1512$_{\rm SW} $\tablenotemark{\dagger} & NGC 1510 \\
\tableline
$\rm M_d$\tablenotemark{a}  &2.4$\pm$0.6$\times$10$^7$ &2.1$\pm$0.4$\times$10$^6$ 
&5.5$\pm$1.3$\times$10$^6$ &5.7$\pm$1.3$\times$10$^6$ &1.7$\pm$3.6$\times$10$^5$\\
$\rm M_{HI, VLA}$\tablenotemark{b}    &4.9$\times$10$^8$   &5.4$\times$10$^6$   
&4.6$\times$10$^7$   &6.2$\times$10$^7$    &4.7$\times$10$^7$ \\
$\rm M_{HI, Parkes}$\tablenotemark{b} &1.7$\times$10$^8$   &1.2$\times$10$^7$   
&1.5$\times$10$^7$   &1.0$\times$10$^7$    &1.4$\times$10$^7$ \\
\tableline
\end{tabular}
%% Any table notes must follow the \end{tabular} command.
\tablenotetext{\dagger}{NGC 1512$_{\rm C}$ represents the central region 
of NGC 1512, while the northeastern/southwestern arms are denoted
by NGC 1512$_{\rm NE}$/NGC 1512$_{\rm SW}$.}
\tablenotetext{^a}{Derived dust mass, in units of M$_{\odot}$, from the model fits of 
\citet{Draine07}.}
\tablenotetext{^b}{Extrapolated {\sc H i} mass, in units of M$_{\odot}$ \citep{Thornley06}. The total {\sc H i} mass is the sum of $\rm M_{HI, VLA}$ and $\rm M_{HI, Parkes}$.}
%\tablecomments{}
\end{center}
\end{table}

\clearpage

\begin{table}
\begin{center}
\caption{Derived Parameters of Two--Blackbody Fit for NGC 1512 and NGC 1510 \label{tab4}}

\begin{tabular}{llllll}
\tableline\tableline
$Parameter$ & NGC 1512 & NGC 1512$_{\rm C}$\tablenotemark{\dagger} 
& NGC 1512$_{\rm NE}$\tablenotemark{\dagger} 
& NGC 1512$_{\rm SW} $\tablenotemark{\dagger} & NGC 1510 \\
\tableline
T$_{\rm w}$\tablenotemark{a} & 24.0     &24.8     &21.4     &20.4     &36.1     \\
T$_{\rm c}$\tablenotemark{a} & 13.8     &16.2     &13.9     &13.8     &15.4     \\
M$_{\rm w}$\tablenotemark{b} & 4.1$\times$10$^6$ &1.7$\times$10$^6$ 
&1.8$\times$10$^6$ &2.3$\times$10$^6$ &4.2$\times$10$^4$ \\
M$_{\rm c}$\tablenotemark{b} & 5.5$\times$10$^7$ &2.5$\times$10$^6$ 
&8.7$\times$10$^6$ &7.1$\times$10$^6$ &1.5$\times$10$^6$ \\
\tableline
\end{tabular}
%% Any table notes must follow the \end{tabular} command.
\tablenotetext{\dagger}{NGC 1512$_{\rm C}$ represents the central region 
of NGC 1512, while the northeastern/southwestern arms are denoted
by NGC 1512$_{\rm NE}$/NGC 1512$_{\rm SW}$.}
\tablenotetext{a}{Best fit values for the temperatures, T$_{\rm w}$ and T$_{\rm c}$ in units of K, from a simple two 
modified blackbody fit, with fixed power-law index of the dust emissivity $\epsilon$=2.}
\tablenotetext{b}{Best fit warm  and cool dust masses, in units of M$_{\odot}$, associated with 
T$_{\rm w}$ and T$_{\rm c}$, respectively.}
%\tablecomments{}
\end{center}
\end{table}

\clearpage

\clearpage

%\begin{center}
%\end{center}

%\begin{figure}
%\figurenum{4}
%    \hspace{-.5mm}
%    \includegraphics[totalheight=3cm,angle=0,origin=c,scale=1.]{1510_sed.ps}%
%    \includegraphics[totalheight=3cm,angle=0,origin=c,scale=1.]{1510_sed2.ps}
%    \hspace{8mm}
%    \incl[totalheight=3cm,angle=0,origin=c,scale=1.]{1512_sed.ps}%
%    \includegraphics[totalheight=3cm,angle=0,origin=c,scale=1.]{1512_sed2.ps}
%    \hspace{8mm}
%    \includegraphics[totalheight=3cm,angle=0,origin=c,scale=1.]{1512core_sed.ps}%
%    \includegraphics[totalheight=3cm,angle=0,origin=c,scale=1.]{1512core_sed2.ps}
%    \hspace{8mm}
%    \includegraphics[totalheight=3cm,angle=0,origin=c,scale=1.]{1512arm1_sed.ps}%
%    \includegraphics[totalheight=3cm,angle=0,origin=c,scale=1.]{1512arm1_sed2.ps}
%    \hspace{8mm}
%    \includegraphics[totalheight=3cm,angle=0,origin=c,scale=1.]{1512arm2_sed.ps}%
%    \includegraphics[totalheight=3cm,angle=0,origin=c,scale=1.]{1512arm2_sed2.ps}
%    \caption{The same SED as shown by Fig. 3, with data points at $\lambda$$>$40$\mu$m 
%fitted by simple modified blackbody spectra. The left panel of each row shows a fit
%by a single blackbody component leaving emissivity index $\epsilon$ a free parameter, 
%while the right one is fitted by two blackbody components, with $\epsilon$ fixed to
%be 2. The best-fit parameters are given in conjunction with the figures.}
%\label{fig4}
%\end{figure}

%% The following command ends your manuscript. LaTeX will ignore any text
%% that appears after it.

\end{document}